\documentclass[%
 reprint,
superscriptaddress,
nofootinbib,
 amsmath,amssymb,
 aps,
 prl,
]{revtex4-2}

\usepackage{graphicx}
\usepackage{dcolumn}
\usepackage{bm}
\usepackage{hyperref}
\usepackage[normalem]{ulem}
\usepackage{xcolor}

\begin{document}

\preprint{APS/123-QED}

\title{Quantum Localization in Geometric Complex Networks}

%%%%%%
%I just uploaded the author information I had in the Master's thesis document. Order to be decided, as well 
%%%%%%

\author{Jan Lewenstein-Sanpera}
\affiliation{Facultat de F\'{\i}sica, Universitat de Barcelona, Mart\'{\i} i Franqu\`{e}s 1, 08028 Barcelona, Spain}

\author{Tobias Grass}
\affiliation{DIPC - Donostia International Physics Center, Paseo Manuel de Lardiz\'abal 4, 20018 San Sebasti\'an, Spain}
\affiliation{IKERBASQUE, Basque Foundation for Science, Plaza Euskadi 5, 48009 Bilbao, Spain}

\author{M. \'Angeles Serrano}
\affiliation{Dpt. Condensed Matter Physics, Facultat de F\'{\i}sica, Universitat de Barcelona, Mart\'{\i} i Franqu\`{e}s 1, 08028 Barcelona, Spain}
\affiliation{Universitat de Barcelona Institute of Complex Systems (UBICS), Barcelona, Spain}
\affiliation{Institució Catalana de Recerca i Estudis Avançats (ICREA), Passeig Lluís Companys 23, E-08010 Barcelona, Spain}

\author{Bruno Juli\'a-Díaz}
\affiliation{Dpt. Quantum Physics and Astrophysics, Facultat de F\'{\i}sica, Universitat de Barcelona, Mart\'{\i} i Franqu\`{e}s 1, 08028 Barcelona, Spain}
\affiliation{
Institut de Ciències del Cosmos, Universitat de Barcelona, 08028 Barcelona, Spain}

\date{\today}

\begin{abstract}
Quantum localization, the phenomenon by which wave diffusion is halted, here understood as the superposition of propagating waves, can have many different origins, such as onsite random potentials in Anderson localization. However, some systems exhibit complex geometries and could potentially reproduce similar results without external potentials. In this work, we study how localization of single particle states is affected by the underlying topology of a simple but highly predictive complex network model: the geometric soft configuration model $\mathbb S^1$. By tuning the model's temperature and heterogeneity parameters, we are able to generate different structural patterns and focus on the transition from a small-world to a non-small-world regime. Our study suggests that small-world networks tend to have extended states, which spread the whole system's size, in contrast to non-small-world networks which tend to localize, showing how complex connectivity can play the role assigned to random energetic disorder.
\end{abstract}

\keywords{Localization, Complex network, Small-world property}
\maketitle

\textit{Introduction} --
Quantum localization transitions have been widely studied in many different systems \cite{Anderson_1958_Localization, Mott_1961_anderslon_1D, Abrahams_1979_scaling_anderson, Jahnke_2008_wave_localization_clustering, Bhattacharjee_2025}. For example, it is known that an onsite random potential causes localization, a phenomenon termed as Anderson localization \cite{Anderson_1958_Localization, Mott_1961_anderslon_1D} which depends on the system's dimensionality and on the noise magnitude \cite{Anderson_1958_Localization, Mott_1961_anderslon_1D}, and has been experimentally validated \cite{Chabe_2008_Ander_loc_expe}. In such systems, localization is a consequence of the superposition of propagating waves that scatter off the introduced random onsite potentials; producing randomly distributed phases \cite{muller_2011_notes_local}. In most of these studies, the geometry is based on regular lattices, such as Euclidean lattices, where the distance between any two points is well defined \cite{Anderson_1958_Localization, Mott_1961_anderslon_1D, Abrahams_1979_scaling_anderson}. Nonetheless, this is not the case for other topologies, such as the ones caused by structural disorder or exotic geometries which require careful consideration \cite{Bhattacharjee_2025, Dutkiewicz_Bruno_BE_conden}. However, many real systems are characterized by a complex underlying topology, well described by complex network architectures \cite{Newman_2018_Networks_BOOK}. This non-trivial connectivity exhibits features such as the \textit{small-world} property: where the shortest path between any two nodes, i.e. the path that connects two nodes and uses the least amount of edges, scales at most logarithmically with the system's size \cite{watts_1998_collective_small_world}. Many complex network models have tried to accurately describe real complex networks \cite{watts_1998_collective_small_world, Barabasi_1999_model, Bollobas_1980_configuration}, but fail to simultaneously exhibit most of the characteristic properties. The $\mathbb S^1$ model \cite{Serrano_2008_SD_complex_networks, Boguna_Serrano_2020_small_world, Serrano_Boguna_2022_NetworkGeometry} was developed to simultaneously generate small-world, heterogeneous, sparse and clustered networks with just tuning the network's parameters. To study quantum-dynamical properties in such complex spaces, random matrix theory can be used \cite{Bohigas_1984_chaos_dyson, Guhr_1998_RMT, Haake_2010_chaos}, which has originally been developed for random interactions \cite{BreitWigner_1936_Slowelectrons}. Through this theory, localized states follow a different eigenvalue distribution than that of extended states, tightly correlated to the system's quantum chaotic behavior \cite{Bohigas_1984_chaos_dyson, Haake_2010_chaos, Zakrzewski_2023} and can be easily studied by numerically decomposing the system's Hamiltonian \cite{Santos_2021_gap_ratio_notes}. 

In this letter, we characterize the quantum localization transition of single particle states over the $\mathbb S^1$ complex network in different connectivity scenarios. We pay special attention to the network's transition between the small-world and the non-small-world regime, and observe how connectivity differences have a strong impact on both the energy distribution as well as the time evolution of single particles on the system. Such topology-driven transitions could be used in the design of programmable quantum simulators and quantum communication networks~\cite{Chen_2026_SWN}, where connectivity could be engineered to control coherent transport~\cite{Dutkiewicz_Bruno_BE_conden, Sachdev_2026_SYK_Hubbard}.\\

\textit{Geometric Complex Network} -- Complex networks are graphs with non-trivial topological features, typically characterized by the simultaneous presence of the small-world property
degree heterogeneity, network sparsity and finite clustering. 
They are used to model real-world networks, such as the Internet or airport connections, and provides a complex underlying topological space \cite{Newman_2018_Networks_BOOK}. However, most complex network models are not able to reproduce all these properties simultaneously from a single connectivity law and are not equipped with an embedded metric space, and this is exactly where the $\mathbb S^D$ geometric model~\cite{Serrano_2008_SD_complex_networks} fits. It generates complex networks which can be tuned to exhibit different configurations, with an isomorphic formulation into the hyperbolic metric space in $\mathbb H^{D+1}$ \cite{Krioukov_Bog_2010_Hyperbolic_graphs}. 

In the $\mathbb S^1$ model \cite{Serrano_2008_SD_complex_networks}, nodes are assigned an angular position $\theta_i$ in a similarity space of $D=1$ represented as a one-dimensional sphere of radius $R$, which determines the distance between any two nodes given by $d_{ij} = R\left(\pi-|\pi-|\theta_i-\theta_j||\right)$. Nodes are also assigned a hidden popularity variable $\kappa_i$, which determines the expected degree and is sampled from a power law distribution $\rho(\kappa)\sim \kappa^{-\gamma}$ with $\gamma$ controlling the degree heterogeneity (e.g., $2<\gamma<3$ are very heterogeneous networks named scale-free, with diverging second moment of the degree distribution). Then, nodes are connected based on the following probability function:
\begin{equation}
    p_{ij} = \frac{1}{1+\left(\frac{d_{ij}}{\mu \kappa_i\kappa_j}\right)^\beta},
\end{equation}
with $\mu$ controlling the average degree of the network and $\beta$ playing the role of an inverse temperature and controlling clustering. This form of the connection probability ensures the maximum entropy principle \cite{Boguna_Serrano_2020_small_world, Serrano_2026_statisticalmechanicsrandomhyperbolic}. By tuning the values of $\beta$ and $\gamma$ the network exhibits different structural transitions \cite{Boguna_Serrano_2020_small_world,Serrano_2026_statisticalmechanicsrandomhyperbolic}, as the geometric to non-geometric transition at $\beta=1$~\cite{van2022anomalous} and the small-world to the non-small-world regime transition. In the small-world regime, the number of nodes grows exponentially fast with the diameter of the network. The transition to the non-small-world regime happens at $\beta=2$ for $\gamma>3$, moving from a highly inter-connected network with long-range connections to a more ``lattice-like" regime dominated by short range connections, while for any $\beta>1$ and $2<\gamma<3$ the model is in an ultra small-world regime where typical distances grow slower than any polynomial of $\ln{N}$. \\

% A summary of these regimes as a function of the parameters $\beta$ and $\gamma$ is given in \autoref{tab: geometric network properties ARTICLE}.

% {\color{violet}\it I think that the table is not needed, but I corrected it in case you want to keep it.}
% \begin{table}[h]
% \caption{ \label{tab: geometric network properties ARTICLE} 
% %Properties 
% {\color{violet}Phases} of the generated $\mathbb S^1$ networks based on the network's parameters. USW = Ultra-small-world{\color{violet}, SW= Small-World,}
% % SWHO = small-world homogeneous, SWHE = small-world heterogeneous
% and NSW = Non-small-world \cite{Serrano_2026_statisticalmechanicsrandomhyperbolic}. We restricted the phase space to the limits used in this article.}
% \begin{ruledtabular}
%     \begin{tabular}{l | r r}
%         \textbf{Domain} & $2<\gamma < 3$ & $\gamma > 3$ \\ 
%         \colrule
%         $1 < \beta < 2$        & USW & 
%         %SWHO
%         {\color{violet}SW}  \\
%         $\beta>2$       & 
%         %SWHE 
%         {USW} 
%         & NSW \\
%     \end{tabular}
% \end{ruledtabular}
% \end{table}

\textit{Single Particle Model} -- Quantum single particle systems can be described through the non-interacting \textit{tight binding Hamiltonian}, which, given a complex network with a set of enumerated nodes $\mathcal N$ and edges $\mathcal E$, is expressed as:

\begin{equation}
    \hat {\mathcal H} = -\sum_{(i,j)\,\in\, \mathcal E} J_{ij}\left(\hat a_i^{\dagger} \hat a_j + \hat a_j^{\dagger} \hat a_i\right) 
    \label{eq: tight binding hamiltonian}
\end{equation}
where $\hat a_i$ and $\hat a_i^{\dagger}$ are the annihilation and creation operators on site $i \in \mathcal N$, and $J_{ij}=J$ is the hopping parameter. Then, in a complex network, $J_{ij}=J \mathcal A_{ij}$, where $\mathcal A_{ij}$ is the adjacency matrix of the underlying graph with entries 1 or 0 when a link is present or absent between the nodes in the corresponding pair. Therefore, it is Hermitian with its set of eigenvalues $\{\epsilon_i\}$ strictly real.

Localization properties can be directly studied through the spacing distribution $P(s)$, defined as the probability distribution of the eigenvalue differences $s_i=\tilde\epsilon_i-\tilde\epsilon_{i+1}$, where $\tilde\epsilon_i$ refers to unfolded spectra with locally normalized eigenvalues \cite{Haake_2010_chaos, Santos_2021_gap_ratio_notes}. The symmetry class of the Hamiltonian characterizes the spacings distribution, which follows a Wigner-Dyson law \cite{Bohigas_1984_chaos_dyson, Haake_2010_chaos}:
\begin{equation}
    P(s)\sim s^{-\theta_D}e^{-s},
\end{equation}
where $\theta_{D}$ is the Dyson parameter. Extended states will be characterized by level repulsion, meaning that eigenvalues are well separated in the spectrum, with the behavior at $s\to0$ with $\theta_D\neq0$ determined by the symmetry class. Because of the time-reversal symmetry of our specific system (with real-valued $J_{ij}$) and that we only consider spinless particles, our Hamiltonian should follow the \textit{Gaussian Orthogonal Ensemble} (GOE) symmetry class which is characterized by $\theta_D = 1$ \cite{Haake_2010_chaos}. However, integrable and often localized systems, are characterized by $\theta_D=0 \longrightarrow P(0)\neq0$, with absence of level repulsion \cite{Haake_2010_chaos, Zakrzewski_2023, Anderson_1958_Localization}, and the exact behavior in our system will be given by the different connectivity regimes.  In addition, these distributions are further characterized by the gap ratio:
\begin{equation}
    \langle \bar r\rangle = \left \langle\min \left(\frac{\epsilon_i-\epsilon_{i-1}}{\epsilon_{i+1}-\epsilon_{i}},\;\frac{\epsilon_{i+1}-\epsilon_{i}}{\epsilon_{i}-\epsilon_{i-1}}\right)\right \rangle,
\end{equation}
which is a cheaper alternative to characterize localization properties.
\begin{figure}[b]
    \centering
    \includegraphics[width=\linewidth]{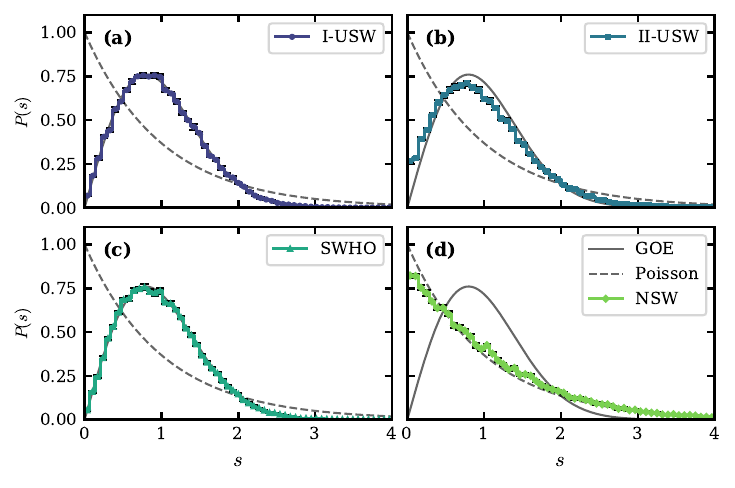}
    \caption{Level spacings distribution $P(s)$ histogram for $\mathbb S^1$ networks with fixed $N=4000$ and $\langle k\rangle=15$. (a) In the I-USW ($\beta=1.21$\;\;$\gamma=2.21$), (b) in the II-USW ($\beta=10.01$\;\;$\gamma=2.21$) and (c) in the SWHO ($\beta=1.21$\;\;$\gamma=8.01$) regimes, where statistics follow a GOE ensemble. (d) In the NSW ($\beta=10.01$\;\;$\gamma=8.01$) regime a Poisson ensemble. Averages and errors are shown for samples of size $m=15$.}
    \label{fig: spacing histogram ARTICLE}
\end{figure}

\begin{figure*}[t]
    \centering
    \includegraphics[width=0.98\linewidth]{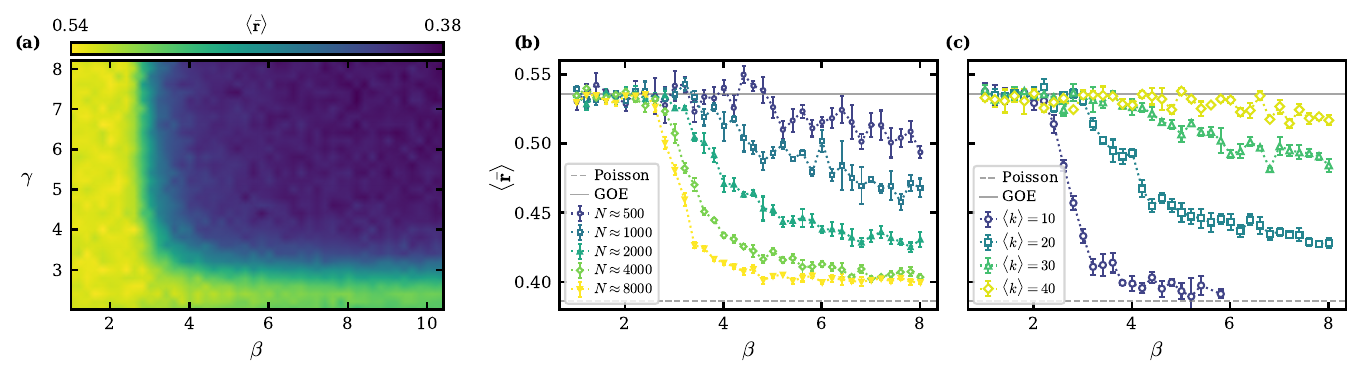}
    \caption{Gap ratio description in the parameter regime. (a) Heat-map representing a possible phase diagram of the gap ratio for $N=4000$ and $\langle k \rangle = 15$ networks where each point is averaged over $m=4$ samples. Transitions from the GOE distribution to a Poisson distribution for the gap ratio for (b) size $N$ dependence with network parameters: $\langle k \rangle = 15$ and $\gamma=6.01$, and (c) for the average degree $\langle k \rangle$ with parameters $N=2000$ and $\gamma=6.01$. (values are averaged over $m=3$ samples). The transition occurs near the small-world to the non-small-world structural transition.}
    \label{fig: combined gap ratio ARRTICLE}
\end{figure*}

Furthermore, localization properties can be studied through the time evolution of single particle states $|\psi(t)\rangle$. The solution of the time-dependent Schrödinger equation is given by: 
\begin{equation}
    |\psi(t)\rangle = \sum_{i=1}^{N} e^{-i\,\epsilon_i\,(t-t_0)/\hbar} \;\;|\hat n_i\rangle \langle \hat n_i|\psi(t_0)\rangle.
    \label{eq: time evolution quantum}
\end{equation}
Here, $|\hat n_i\rangle$ are the eigenvectors with eigenvalue $\epsilon_i$ \cite{Fehske_2009_quantum_time_approaches}. From this expression, one can characterize localization properties through Shannon's entropy: $S(t) = -\sum_{i\,\in\, \mathcal N}p_i(t) \ln p_i(t)$, where the sum runs over each node in the system and $p_i(t) = 1/T \int_T |\langle\psi_i(t)  |\psi_i(t)\rangle|^2$ is the time averaged probability at every node. Values closer to $0$ will indicate high localization, while a fully extended state should tend to $-\ln 1/N$. In the limit $T\longrightarrow\infty$ we obtain the time-averaged distribution which can be directly computed through the time evolution operators above mentioned. \\

\textit{Localization on Complex Networks} -- In order to test how the small-world property affects localization, we choose a network representative in each of the regimes, with $N=4000$ nodes and an average degree of $\langle k \rangle =15$. The chosen networks were: I-USW with $\beta=1.21$ $\gamma=2.21$; II-USW with $\beta=10.01$ $\gamma=2.21$; SWHO with $\beta=1.21$ $\gamma=8.01$; and NSW with $\beta=10.01$ $\gamma=8.01$, unless otherwise stated. Level spacings distribution show a clear dependence on the small-world property of the underlying network, as can be observed in Fig.~\ref{fig: spacing histogram ARTICLE}. For the previously described networks in Fig.~\ref{fig: spacing histogram ARTICLE}, we observe how the non-small-world regime, defined for $\beta>2$ and $\gamma>3$ follows a Poisson distribution, with a high probability that the eigenvalues are very close one to another as $s\longrightarrow0$ since the maximum is attained at this point. To mitigate finite-size effects, the lowest and highest $5\%$ eigenvalues were discarded and the gap ratio for a sample of $m=15$ was computed. For the NSW network, the gap ratio took a value of $\langle r \rangle_{\rm NSW} =0.3991 \pm 0.0012$ which lays close to the theoretical value $\langle r \rangle_{\rm Poisson} \approx0.3863$ for random matrix ensembles which belong to the Poisson ensemble \cite{Atas_2013_Gap_ratio_values}. This is characteristic of integrable systems which are likely to localize. On the contrary, we observe how all the small-world regimes exhibit a strong level repulsion as $s\longrightarrow0$ and they closely follow the GOE distribution, as expected by the real Hermitian Hamiltonian and time-reversal. In this case, the gap ratios were computed in the same fashion as previously described and we obtained $\langle r \rangle_{\rm I-USW} =0.5311 \pm 0.0009$, $\langle r \rangle_{\rm II-USW} =0.5281 \pm 0.0013$, and $\langle r \rangle_{\rm SWHO} =0.5327 \pm 0.0011$. All of them lay in close agreement to the theoretical value $\langle r \rangle_{\rm GOE} \approx 0.53590$ \cite{Atas_2013_Gap_ratio_values} of GOE ensembles, usually described through quantum chaotic behavior and not localizing.

To characterize the regime changes we explored a phase diagram for the gap ratio for network parameters ranging between $\beta\in(1,10)$ and $\gamma\in(2,8)$, fixing the network's size to $N=4000$ and $\langle k \rangle =15$, which can be observed in Fig.~\ref{fig: combined gap ratio ARRTICLE}. This heat diagram clearly shows a sharp gap ratio transition from the characteristic GOE to the Poisson ensemble description at the small-world transition. However, we must note that the transition due to the values of $\beta$, which control network's clustering, happens at a slightly larger value, with $\beta\approx3$. As we can also observe in Fig.~\ref{fig: combined gap ratio ARRTICLE} (b) and (c), this is caused by finite effects and sparsity in the network generation algorithms, which both directly modify the sharpness and location of the gap ratio transition. Indeed, we observe how increasing network sizes produce sharper transitions and faster convergence to the expected Poisson gap ratio. However, in the small-world regime (with $\beta<2$), the gap ratios are homogeneous across sizes and therefore independent of the network's size. On the other hand, sparsity, which is controlled by the average degree $\langle k \rangle$, not only sharpens the transition but it also moves the jump closer to the small-world transition values. 

\begin{figure}[t]
    \centering
    \includegraphics[width=0.99\linewidth]{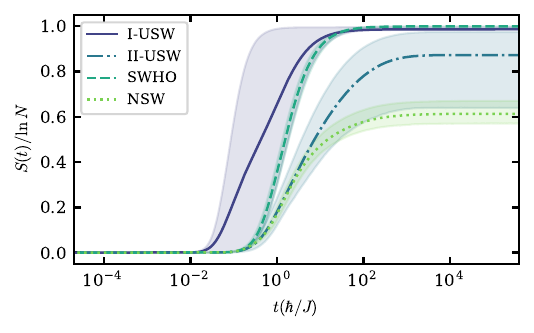}
    \caption{Entropy evolution across different networks with different realizations ($k=5$) and $500$ nodes as different initial states. 
    I-USW ($\beta=1.21$\;\;$\gamma=2.21$),  and II-USW ($\beta=10.01$\;\;$\gamma=2.21$) and SWHO ($\beta=1.21$\;\;$\gamma=8.01$) networks tend to have a larger entropy, indicating that the time-averaged state spans most of the system (extended), while the NSW network ($\beta=10.01,\;\;\gamma = 8.01$) reaches smaller values, $\sim2$-fold smaller, and indicating larger localization. The cloud represents the $[15, 85]$ percentile.}
    \label{fig: entropy evolution regimes ARTICLE}
\end{figure}

We further study localization through Shannon's entropy, which can be observed for the four different representatives in Fig.~\ref{fig: entropy evolution regimes ARTICLE}. Again, we note how the NSW regime is characterized by more localized states in contrast to their small-world counterparts, with the NSW taking a maximum at almost half the value of small-world regimes, meaning that the time-averaged distribution in small-world networks spans most of the system, generally independent of the initial position. However, II-USW networks seem to have a higher variability, which is caused by highly disconnected nodes (leaves) and to lesser extent by hubs, since eigenvalues tend to localize on these nodes and therefore particles starting there have a higher chance of localizing there. Interestingly as well, we observe how the time a single particle needs to spread across the system is generally larger in non-small-world networks due to the smaller amount of edges between nodes.

\begin{figure*}
    \centering
    \includegraphics[width=0.99\linewidth]{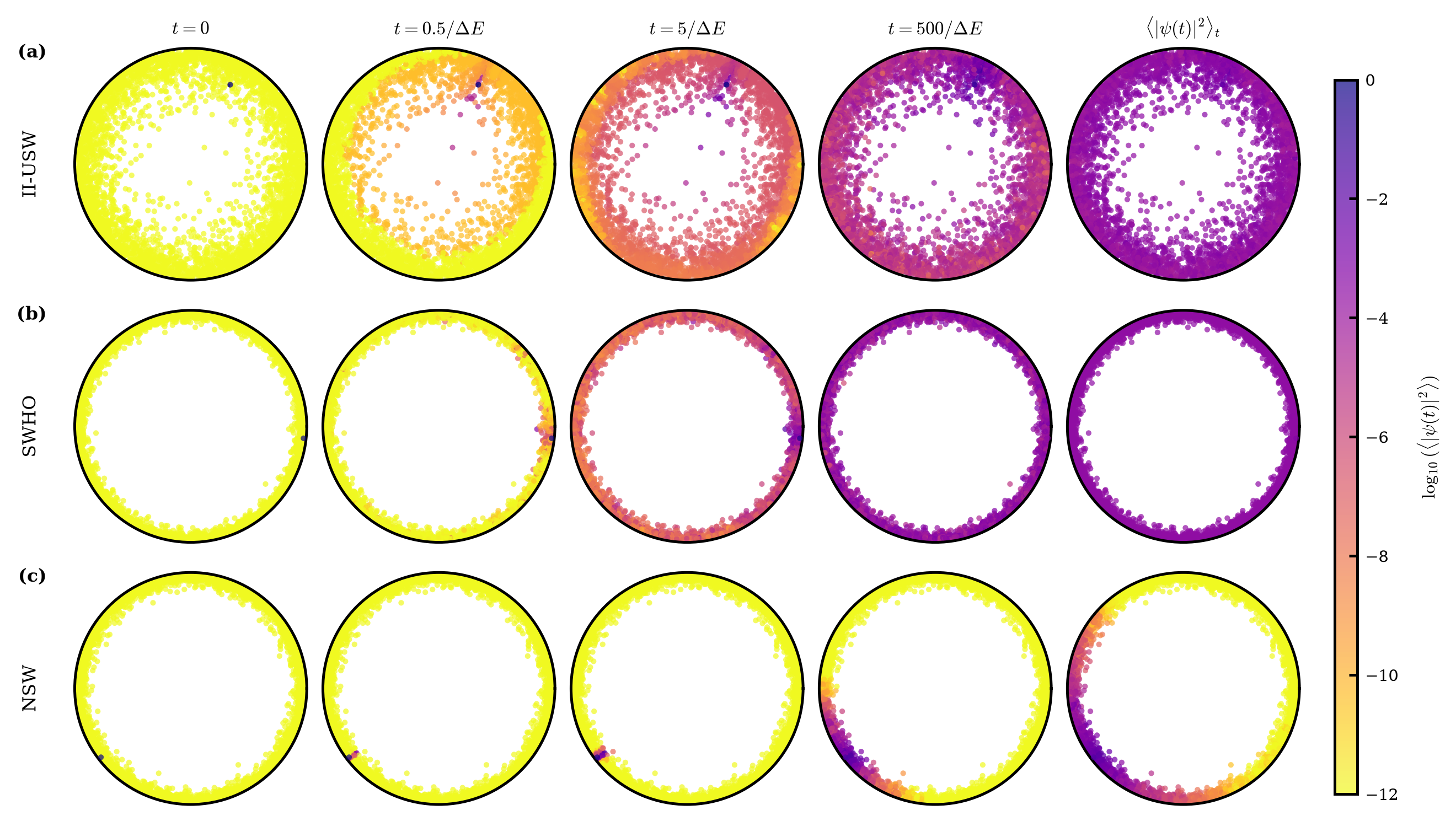}
    \caption{Time evolution of different states in different regimes, with probabilities plotted in logarithmic scale. To ease the representation, links have not been drawn and we follow the $\mathbb H^2$ formalism to plot represent $\mathbb S^1$ networks. (a) II-USW network with $\beta=10.01$ and $\gamma=2.21$ which has a bandwidth $\Delta E = 136.7175$; (b) SWHO network with $\beta=1.21$ and $\gamma=8.01$ with $\Delta E=24.7345$; and (c) in a NSW homogeneous network with $\beta=10.01$ and $\gamma=8.01$ with $\Delta E=28.7145$. All networks have $N=4000$ nodes and average degree $\langle k \rangle=15$, with a single particle starting at a node of degree $15$. Dimensionless time $t$ in units of $\tau = \hbar/J$. The value $\langle |\psi(t)|^2\rangle_t$ refers to the time-averaged probability as $T\to\infty$.}
    \label{fig: time evolution regimes ARTICLE}
\end{figure*}

On this matter, example runs for the previously described representatives with a single particle starting at a random node of degree $k=15$ are plotted in Fig.~\ref{fig: time evolution regimes ARTICLE}. There we have used the $\mathbb H^2$ formalism to represent the $\mathbb S^1$ networks in the hyperbolic plane, whereby the hidden popularity variables become radial coordinates with hubs closer to the center of the disk. Even if they represent a single network realization and evolution, with little statistical significance, it helps to grasp the different localization behaviors due to the small-world nature of the underlying complex network, with the NSW example clearly showing a localization behavior. In contrast, both the II-USW and SWHO networks show no localization at the time-averaged state, with the particle extending all over the network in a short time-scale. On this matter, we note that different time scales determine the particle's spreading across the network, with a characteristic time that is controlled by the spectrum's bandwidth: $\Delta E = \epsilon_{\max} - \epsilon_{\min}$. This enables us to observe how drastically different timescales shape the particle's localization, a feature that can also be observed in Fig.~\ref{fig: entropy evolution regimes ARTICLE}. \\

\textit{Discussion} -- Small-world complex networks are fundamental 
representations of real-world complex systems. 
Our study further supports the idea that quantum localization is highly dependent on the underlying topological space \cite{Jahnke_2008_wave_localization_clustering, Kantelhardt_wave_localisation_2010, Bhattacharjee_2025}, and directly affected by the small-world regime of the studied networks \cite{Zhu_2000_local_delocal_small_world, Chen_2026_SWN}. Because of the plasticity of $\mathbb S^1$ complex network models, we were able to parametrize the network's phase transition and observe that quantum systems on topologies with the small-world property behave in a chaotic manner and have a higher fraction of non-localized states compared to non-small-world complex networks, which are characterized by localized single particle states with eigenvalue level spacings distributions following a Poisson distribution. This enhances the idea that single-particle systems can undergo a metal-insulator transition induced by the underlying topological complex network \cite{Bhattacharjee_2025}, without relying on the quantum interactions such as in the Bose-Hubbard model \cite{Halu_2012_Bose-HUbbard_network, Dutkiewicz_Bruno_BE_conden} or on on-site potentials which create an energy barrier as in Anderson's model \cite{Anderson_1958_Localization, Sade_2005_Anderson_spectral_stat}, where particle states become trapped due to the topological constraints.

However, the effect is not purely due to the small-world structural transition: degree heterogeneity, sparsity and clustering play a significant role and has been observed to produce localization \cite{Jahnke_2008_wave_localization_clustering, Tikhonov_2021_local_in_randoM_graphs}. These properties can directly modify the behavior of single particle states, specially if particles are placed on leaves (very low degree nodes) or hubs (very high degree nodes) which directly relate to the spectral edges of the Hamiltonian \cite{Newman_2018_Networks_BOOK, Garcia_Perez_2018_eigenvalues_S1}.

In conclusion, the quantum behavior of single-particle systems on the geometric complex network $\mathbb S^1$ model  is strongly correlated to the underlying topological space, with localization highly dependent on them. These networks, which are general and flexible, are ideal to study different structural regimes and, in contrast to regular lattices with external potentials, their structural properties determine the quantum behavior of the system. 

\begin{acknowledgments}
M.\'A.S acknowledge support from Grant No. PID2022-137505NB-C22 funded by MCIN/AEI/10.13039/501100011033 and by ERDF/EU and Grant No. PID2025-169261NB-C21 funded by MICIU/AEI/10.13039/501100011033. B.J-D acknowledges support from PID2023-147475NB-I00 funded by MICIU/AEI/10.13039/501100011033 and FEDER, UE, by grants 2021SGR01095 from Generalitat de Catalunya, and by Project No. CEX2019-000918-M of ICCUB (Unidad de Excelencia Mar´ıa de Maeztu).
T.G. acknowledges funding by the Department of Education of the Basque Government through the IKUR Strategy under the collaboration agreement between the Ikerbasque Foundation and DIPC, and by the Agencia Estatal de Investigación (AEI) through Proyectos de Generación de Conocimiento PID2022-142308NA-I00 (EXQUSMI) and PID2025-175305NB-I00 (AQUASFONDAS).
 \end{acknowledgments}

\appendix
%\nocite{*}

\bibliography{apssamp}

\end{document}